\documentclass[reprint,superscriptaddress,aps,prb,floatfix]{revtex4-2}

\usepackage{mathtools}
\usepackage{graphicx}
\usepackage{hyperref}
\usepackage[export]{adjustbox}
\usepackage{tikz}
\usepackage{braket}
\usepackage{textcomp,gensymb}
\usepackage{dsfont}
\usepackage{algpseudocode}
\usepackage{bbold}
\usepackage{changes}
\usepackage{comment}
\makeatletter
\def\parsecomma#1,#2\endparsecomma{\def\page@x{#1}\def\page@y{#2}}
\tikzdeclarecoordinatesystem{page}{
    \parsecomma#1\endparsecomma
    \pgfpointanchor{current page}{north east}
    \pgf@xc=\pgf@x%
    \pgf@yc=\pgf@y%
    \pgfpointanchor{current page}{south west}
    \pgf@xb=\pgf@x%
    \pgf@yb=\pgf@y%
    \pgfmathparse{(\pgf@xc-\pgf@xb)/2.*\page@x+(\pgf@xc+\pgf@xb)/2.}
    \expandafter\pgf@x\expandafter=\pgfmathresult pt
    \pgfmathparse{(\pgf@yc-\pgf@yb)/2.*\page@y+(\pgf@yc+\pgf@yb)/2.}
    \expandafter\pgf@y\expandafter=\pgfmathresult pt
}
\makeatother

\newcommand{\fref}[1]{Fig.~\ref{#1}}

\usepackage{xcolor}

\begin{document}

\preprint{APS/123-QED}

\title{Competition between \texorpdfstring{$d$}{d}-wave superconductivity and magnetism \\
in uniaxially strained \texorpdfstring{Sr$_2$RuO$_4$}{Sr2RuO4}}

\author{Jonas B.~Profe}
\affiliation{Institute for Theoretical Solid State Physics,
    RWTH Aachen University and JARA Fundamentals of Future Information Technology, 52062 Aachen, Germany}
\affiliation{Center for Computational Quantum Physics, Flatiron Institute, 162 Fifth Avenue, New York, New York 10010, USA}
\author{Sophie Beck}
\affiliation{Center for Computational Quantum Physics, Flatiron Institute, 162 Fifth Avenue, New York, New York 10010, USA}
\author{Dante M.~Kennes}
\affiliation{Institute for Theoretical Solid State Physics,
    RWTH Aachen University and JARA Fundamentals of Future Information Technology, 52062 Aachen, Germany}
\affiliation{Max Planck Institute for the Structure and Dynamics of Matter and Center Free-Electron Laser Science, Hamburg, Germany}
\author{Antoine Georges}
\affiliation{Coll\`ege de France, Universit\'e PSL, 11 place Marcelin Berthelot, 75005 Paris, France}
\affiliation{Center for Computational Quantum Physics, Flatiron Institute, 162 Fifth Avenue, New York, New York 10010, USA}
\affiliation{Centre de Physique Théorique, Ecole Polytechnique, CNRS, Institut Polytechnique de Paris, 91128 Palaiseau Cedex, France}
\affiliation{DQMP, Université de Genève, 24 quai Ernest Ansermet, CH-1211 Genève, Suisse}

\author{Olivier Gingras}
\affiliation{Center for Computational Quantum Physics, Flatiron Institute, 162 Fifth Avenue, New York, New York 10010, USA}

\date{\today}
\begin{abstract}
The pairing symmetry of Sr$_2$RuO$_4$ is a long-standing fundamental question in the physics 
of superconducting materials with strong electronic correlations. 
We use the functional renormalization group to investigate the behavior of  
superconductivity under uniaxial strain in a two-dimensional realistic model of Sr$_2$RuO$_4$ 
obtained with density functional theory and incorporating the effect of spin-orbit coupling. 
We find a dominant $d_{\rm{x}^2-\rm{y}^2}$ superconductor mostly hosted by the $d_{\rm{xy}}$-orbital, with no other closely competing superconducting state.
Within this framework we reproduce the experimentally observed enhancement of the critical temperature under strain and propose a simple mechanism driven by the density of states to explain our findings.
We also investigate the competition between superconductivity and spin-density wave ordering as a function of interaction strength.
By comparing theory and experiment, we discuss constraints on a possible degenerate partner of the $d_{\rm{x}^2-\rm{y}^2}$ superconducting state.
\end{abstract}

\maketitle

\section*{Introduction}
Almost 30 years after the discovery of superconductivity in Sr$_2$RuO$_4$ (SRO)~\cite{Maeno1994}, the symmetry of its superconducting order parameter (SCOP) remains an open question.
Initially, its similarities with $^3$Helium made it a prime candidate for spin-triplet pairing~\cite{rice1995sr2ruo4}, corroborated by various measurements~\cite{ishida1998spin,NMR_spin_poli_ishida, Neutron_scattering_duffy,PhysRevB.66.064522, liu2003tunneling,nelson2004odd,liu2010phase, PhysRevLett.92.047002, PhysRevLett.86.5986}.
Along with observations of time-reversal symmetry breaking (TRSB) supporting a two-component order parameter~\cite{PhysRevLett.97.167002, luke1998time}, the superconducting (SC) state was believed for a long time to be a chiral $\rm{p}$-wave spin-triplet. 
However, conflicting evidence presented in various studies remained to be explained~\cite{RevModPhys.75.657,review_maeno2012}.
First, the presence of nodal excitations is unexpected in a chiral $\rm{p}$-wave SC~\cite{PhysRevLett.86.2653, PhysRevX.7.011032, PhysRevLett.86.2649, doi:10.1143/JPSJ.69.572}.
Second, the low critical field $H_{c2}$ exhibited by SRO is typical for Pauli-limited spin-singlet SC~\cite{PhysRevB.107.064509} and the transition into the normal state upon applying a magnetic field appears to be first-order~\cite{PhysRevB.90.220502, doi:10.7566/JPSJ.83.083706}, with indications of a Fulde-Ferrell-Larkin-Ovchinnikov state for a certain parameter range, strongly pointing to a singlet SCOP~\cite{kinjo_superconducting_2022}.
Third, none of the topologically protected edge states predicted in chiral $\rm{p}$-wave states were observed in experiments~\cite{PhysRevB.76.014526, PhysRevB.81.214501, PhysRevB.89.144504, Kreisel_2021}, although this could be explained by a high Chern number~\cite{PhysRevLett.115.087003}.

In recent years, the chiral $\rm{p}$-wave picture has basically been dismissed.
First, the careful replications of key nuclear magnetic resonance experiments previously interpreted as supporting spin-triplet pairing
have highlighted a heating effect and instead concluded that the SCOP corresponds to spin-singlet 
pairs~\cite{pustogow2019constraints, ishida2020reduction, PhysRevLett.125.217004}.
Second, applying uniaxial strain along the $\rm{x}$ principal crystallographic axis was found to enhance the critical temperature ($T_{\rm{c}}$)~\cite{PhysRevB.98.094521, steppke2017strong} and to lower the Fermi liquid coherence scale~\cite{Chronister_2022}.
This enhancement was shown to be maximal where the FS undergoes a Lifshitz transition, corresponding to a van Hove singularity (vHs) in the density of states (DOS), at a time-reversal invariant momentum point, inconsistent with odd-parity SCOPs like $\rm{p}$-wave~\cite{sunko_direct_2019}.
Nowadays, a consensus appears to be crystallizing around the spin-singlet and even-parity natures of the SCOP, yet the debate is still ongoing.
While ultrasounds experiments support the conclusion of a two-component order parameter~\cite{Benhabib2021, Ghosh2021} inferred by the observation of TRSB and the splitting between $T_{\rm{c}}$ and the TRSB transition temperature~\cite{grinenko_split_2021, PhysRevB.107.024508}, there are no two-temperature signature in bulk thermodynamical experiments such as specific heat and elastocalorimetry as well as scanning SQUID microscopy~\cite{doi:10.1073/pnas.2020492118, li2022elastocaloric, palle2023constraints,mueller2023constraints}. 
As a result of this plethora of experimental evidence, SRO can be seen both as a critical playground for testing new theories with the goal of potentially unifying some of these contradicting observations and as a testbed to verify whether our interpretations of specific experiments are valid.
Either way, it constitutes an ideal system to considerably advance our understanding of the mechanisms for unconventional superconductivity~\cite{mackenzie2017even}.

Many theoretical proposals have been put forward as potential SCOPs.
Initially classified as a chiral $\rm{p}$-wave~\cite{PhysRevLett.105.136401,wang2013theory,PhysRevB.91.155103,PhysRevLett.122.027002,PhysRevB.97.060510,scaffidi_pairing_2014,PhysRevB.94.104501,Acharya2019}, the recent experimental evidence motivated further proposals, including two-dimensional states such as $s+id_{\rm{xy}}$~\cite{romer_superconducting_2021,PhysRevResearch.4.033011,moon2023effects,PhysRevB.106.174518}, $d_{\rm{x}^2-\rm{y}^2}+ig_{\rm{xy}(\rm{x}^2-\rm{y}^2)}$~\cite{Kivelson_2020, PhysRevB.106.054516,PhysRevB.106.134512,PhysRevB.106.174518, PhysRevB.104.054518, PhysRevB.108.014502}, a combination of even and odd-parity irreducible representations (irreps)~\cite{PhysRevB.107.014505}, inter-orbital pairing~\cite{Kaba_2019, PhysRevB.106.214520}, $d+d$~\cite{Nica_2021}, and $d_{\rm{x}^2-\rm{y}^2}$ (plus odd-frequency)~\cite{PhysRevLett.107.277003, gingras_superconducting_2019, gingras_superconductivity_2022, gingras2023frequencydependent}.
Some three-dimensional states were also proposed, for example E$_g$ $d_{\rm{yz}}+id_{\rm{zx}}$~\cite{suh_stabilizing_2020,PhysRevResearch.4.023060} and helical $p_{\rm{x}}+p_{\rm{y}}$~\cite{huang_possible_2018}.

A general overview of possible ordering states in terms of their irreducible representations 
is given in Ref.~\citenum{Ramires_micros_perspective_2019, Kaba_2019}.

In this paper, we investigate the leading superconducting instabilities of SRO using functional renormalization group (FRG) calculations~\cite{metzner_functional_2012}, applied to a 
realistic model of the electronic structure derived from density functional theory (DFT)~\cite{PhysRev.140.A1133} that includes spin-orbit coupling (SOC).
Note that previous studies of SRO using FRG were performed on tight-binding models, fitted to photoemission spectroscopy measurements~\cite{wang2013theory,PhysRevLett.122.027002,Liu_2017,PhysRevB.101.064507,PhysRevB.97.224522,PhysRevB.106.134512}.
In order to compare to experiments, we study the effect of uniaxial strain, tracking the evolution of $T_{\rm{c}}$ as well as the type of ordering.
We find a phase diagram with two different magnetic orders that compete with a single SCOP transforming like the B$_{\rm{1g}}$ irrep (often labelled as $d_{\rm{x}^2-\rm{y}^2}$-wave). 
This competition is found to depend sensitively on the choice of interaction parameters.  
We show that a proper range of parameters lead to an 
increase of the superconducting $T_{\rm{c}}$ in good agreement with experiments.   

\section*{\label{sec:results}Results}
\paragraph*{Electronic structure. ---}
To describe the low energy electronic structure of SRO for the different strain values, we perform \textit{ab initio} DFT calculations downfolded onto the $t_{2g}$ orbitals of the ruthenium atoms using maximally localized Wannier functions as detailed in the methods section.
There is an extensive literature using tight-binding Hamiltonians from ab-initio electronic structure calculations based on DFT for SRO, presenting overall very consistent results at the DFT level.
Beyond DFT, it was shown that replacing the local SOC parameter of $100$~meV obtained withing DFT with a value twice as large 
$\lambda_{\text{SOC}}\simeq 200$~meV yielded a considerably better agreement with the 
experimentally observed Fermi surface~\cite{tamai_high-resolution_2019}, in consistent with the correlation-induced 
enhancement predicted theoretically~\cite{PhysRevLett.116.106402,kim_SOC_2018}.
Indeed, this  parameter adjustment leads to improved agreement with the FS both in the unstrained case and with the experimentally observed critical strain~\cite{tamai_high-resolution_2019, PhysRevLett.112.127002, PhysRevLett.101.026406,PhysRevB.74.035115, PhysRevLett.116.106402,kim_SOC_2018, PhysRevLett.85.5194, PhysRevB.100.245139}.
A comparison to a selection of previously published tight-binding models is shown in App.~\ref{app:dft}.
Note that the addition of SOC breaks the $SU(2)$ spin symmetry, but preserves an orbitally dependent $SU(2)$ (so called pseudospin) symmetry~\cite{gingras_superconductivity_2022}. We keep the SOC fixed for all strain values~\footnote{Unpublished DMFT data shows that the corrections of the SOC due to strain are negligible.}.

In order to account for strong electronic correlations in this multi-orbital system, we use for most parts of this paper the $O(3)$ symmetric 
Hubbard-Kanamori parametrization of the interaction Hamiltonian ~\cite{kanamori1963electron}, which involves two key energy scales: the on-site Hubbard repulsion $U$ and the Hund coupling $J$ - see methods. 
As done routinely in FRG calculations~\cite{PhysRevLett.122.027002, Liu_2017, PhysRevB.101.064507, PhysRevB.97.224522, scherer2022chiral,Klebl_2023,PhysRevB.106.125141, wang2013theory,PhysRevLett.111.097001, PhysRevLett.110.126405, PhysRevB.107.125115, PhysRevB.86.020507, Hauck_2021, PhysRevB.102.195108, PhysRevB.102.085109, PhysRevB.96.205155, PhysRevB.99.245140}, 
we neglect the flow of the self-energy in our calculations. 
Hence, the interaction parameters $(U,J)$ should be considered as effective 
interactions with significance within our FRG framework rather than 
having a first-principle meaning. In this perspective, it is 
important to explore how the various instabilities are tuned 
by varying the interaction parameters.
The guiding principles for constructing the model utilized follow two rules
\begin{itemize}
    \item We start from state of the art DFT with added spin-orbit coupling $\lambda_{\text{SOC}}$ such that we do have the right Fermi surface.
    \item The choice of our interaction parameters is restricted in such a manner that, from the Wannier model defined above, we recover the correct peak positions of the interacting spin-spin susceptibility as measured in experiments.
\end{itemize}

\begin{figure}
    \centering
    \includegraphics[width = \columnwidth]{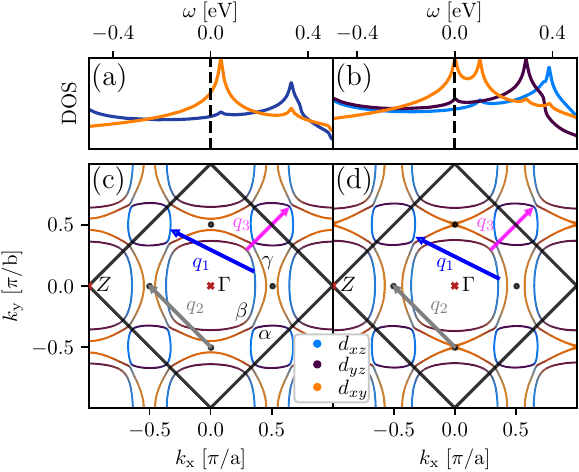}
    \caption{
    \textbf{Partial density of states and Fermi surface for the three $t_{2g}$ orbitals} of the unstrained (left) and the optimally strained (right) systems in the upper anf lowe panels respectively.
    The optimal stain of $0.8\%$ corresponds to the system being closest to the Lifschitz transition.
    Here, $a$ ($b$) is the lattice parameter in the $x$ ($y$) direction, with $a = b$ in the $\epsilon_{\rm{xx}} = 0$ case.
    The black dots indicates the position of the vHs of the $d_{\rm{xy}}$ orbital.
    The three dominant spin-density wave ordering vectors $q_{\rm{1}}$, $q_{\rm{2}}$ and $q_{\rm{3}}$ are highlighted in black, gray and pink, respectively.
    The first Brillouin zone is marked by a black square. We mark the $\Gamma$ and $Z$ point by red crosses. Furthermore, we labelled the $\alpha$, $\beta$ and $\gamma$ sheets on the FS for $\epsilon_{\rm{xx}} = 0$.  In the unstrained case, the partial DOS for the $d_{\rm{xz}}$ and $d_{\rm{yz}}$ orbitals is identical, which is emphasized by the color mixing.}
    \label{fig:FS_and_DOS}
\end{figure}

The FS and the density of states (DOS) obtained 
from this downfolded $t_{2g}$ 
model are displayed on Fig.~\ref{fig:FS_and_DOS}.
The left column corresponds to the unstrained system ($\epsilon_{\rm{xx}}=0$) 
and the right columns to the optimally uniaxially strained system
for which the Fermi level is at the 
vHs ($\epsilon_{\rm{xx}}=\epsilon_{\rm{xx}}^{\text{vHs}}$).
Note that $\epsilon_{\rm{xx}}^{\text{vHs}}$ does not include quasi-particle renormalization and therefore is not the same value as in experiments.
The D$_{\rm{4h}}$ space group symmetry of the unstrained system is 
lowered down to D$_{\rm{2h}}$ by uniaxial strain and the B$_{\rm{1g}}$ irreducible representation of D$_{\rm{4h}}$, of greatest relevance to our study, turns into the A$_{\rm{g}}$ irreducible representation of D$_{\rm{2h}}$.

Note the slightly unusual presentation of the FS in Fig.~\ref{fig:FS_and_DOS}:  this is due to the transformation from a tetragonal basis into a $x$-$y$ plane which has to be done in this fashion to ensure periodicity of the 
downfolded model in the two-dimensional primitive cell.
Due to this, we have not a single but two $k_z$ values in the first primitive cell, i.e. the $Z$-point is located at the corner of the black square. All the results that follow are obtained for the quasi-2D model restricted to $k_z = 0$ and $k_z=2\pi/c$, shown on Fig.~\ref{fig:FS_and_DOS}. We verify that they are agreeing with the results from full 3D calculations in App.~\ref{app:3D}.

The lowering of the symmetry under uniaxial strain lifts the degeneracy between the $d_{\rm{xz}}$ and the $d_{\rm{yz}}$ orbitals of the ruthenium atom, as seen in the DOS in \fref{fig:FS_and_DOS}.
It also splits the $d_{\rm{xy}}$ van-Hove singularity into two parts: one drifting away from the FS ($x$-direction) and one drifting towards the FS and crosses it at the Lifschitz transition ($\epsilon_{\rm{xx}}^{\text{vHs}} \sim 0.8\%$ strain).
On the FS shown in \fref{fig:FS_and_DOS}, we also highlight the dominant nesting vectors of the bare particle-hole susceptibility ($\chi^0_{\text{PH}}$), see App.~\ref{app:ph_susz}.
First, $q_1=(2\pi(3a)^{-1}, \pi(3b)^{-1})$ (and all those related by symmetry) connects the $\alpha$ and $\beta$ sheets of the FS.
Second, $q_2 = (\pi(2a)^{-1}, \pi(2b)^{-1})$ is connecting two van-Hove singularities and should become relevant at large interactions.
Third, $q_3 = (\pi(3a)^{-1}, \pi(3a)^{-1})$ also connects the $\alpha$ and $\beta$ sheet of the FS.
These vectors are consistent with the dominant spin fluctuations observed in neutron scattering experiments~\cite{PhysRevB.84.060402, PhysRevLett.122.047004}. Note that there is a family of nesting vectors connecting $\alpha$ and $\beta$ sheet, all close to $q_{1}$.

\begin{figure}[t]
    \centering
    \includegraphics[width = \columnwidth]{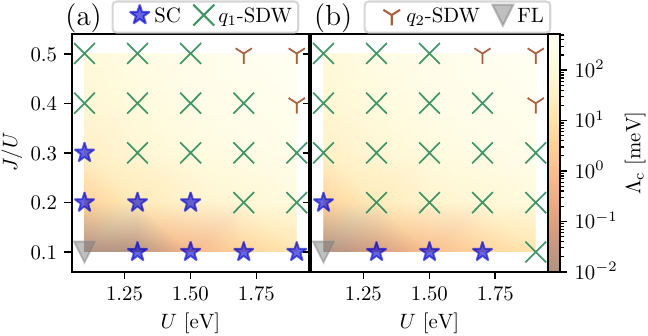}
    \caption{
    \textbf{Phase diagrams in the $U$-$J$ parameter space} for the unstrained (a) and the optimally strained (b) systems.
    The background color indicates the critical scale $\Lambda_c$, proportional to the ordering energy scale, with the corresponding color bar on the right.
    The phases are either a B$_{\rm{1g}}$ superconductor (A$_{\rm{g}}$ under uniaxial strain), two different spin-density waves (SDWs) or a Fermi-liquid (FL).
    The $q$-vectors associated to the SDWs are shown in Fig.~\ref{fig:FS_and_DOS}.
    }
    \label{fig:phase_diag}
\end{figure}

With these insights from the non-interacting FS and DOS in mind, we proceed with the phase diagrams as a function of $U$ and $J$ for the unstrained and the $\epsilon_{\rm{xx}} = \epsilon_{\rm{xx}}^{\text{vHs}}$ cases.
The results are presented on \fref{fig:phase_diag}.
The background color corresponds to the energy scale $\Lambda_c$ 
(expressed in meV)  
at which a divergence of the corresponding coupling is observed. 
The fastest divergent coupling corresponds to the dominant instability, 
which can either be superconductivity (in which case $\Lambda_c$ is 
expected to be proportional to the Berezinskii–Kosterlitz–Thouless~\cite{berezinskii1971destruction, berezinskii1972destruction, Kosterlitz_1973} critical temperature $T_{\text{BKT}}$) or 
a spin-density wave (SDW) (in which case $\Lambda_c$ can be interpreted as the characteristic scale associated with the growth of the correlation length)
~\footnote{Because of Mermin-Wagner theorem (which is not obeyed by FRG), 
a finite $T_{\rm{c}}$ is expected for the BKT transition into the SC phase, 
while $T_{\rm{c}}=0$ for an SDW breaking $SU(2)$.}

At the lowest $U$ and $J$ values, we find no divergence down to the lowest energy scale resolvable with our momentum resolution, and thus 
conclude that the system remains in the Fermi liquid (FL) state down to that scale.  
Apart from this unique point, we find three types of instabilities. 
Up to moderate $U$ and low but finite $J$, we find a 
d$_{x^2-y^2}$ superconducting instability (corresponding to  
B$_{\rm{1g}}$ symmetry for the unstrained system, turning into A$_{\rm{g}}$ for the strained one).
Upon increasing $U$ or $J$, we find that the dominant instability becomes a SDW with ordering vector $q_1$.
At even larger $U$ and $J$, the system undergoes a high-temperature transition to another SDW phase characterized by the ordering vector $q_2$.
These ordering vectors are visible in both non-interacting and interacting susceptibilities.
Since we do not incorporate the effect of the self-energy, we cannot observe the shift of the $q_3$ peak observed in Ref.~\cite{moon2023effects}.

\paragraph*{Phase diagram and magnetic orderings. ---}

The $q_1$-SDW is driven by strong nesting between the $\alpha$ and $\beta$ sheets.
A corresponding peak in the spin-spin susceptibility has been well discussed both in the context of experimental observations~\cite{doi:10.1143/JPSJ.81.124710, PhysRevB.66.064522} and theoretical discussions~\cite{gingras_superconducting_2019,PhysRevResearch.4.033011}.
It should be noted that this vector is connecting two different values of $k_z$ when backfolded in the three dimensional Brillouin zone.
Its in plane analog, $q_3 = (\pi(3a)^{-1}, \pi(3a)^{-1})$, was found to be subleading in earlier three-dimensional studies using the random phase approximation (RPA)~\cite{PhysRevResearch.4.033011}.
The $q_3$ peak is also found in DMFT calculations including vertex corrections~\cite{Strand_2019, moon2023effects}.
Here, we find the $q_1$ ordering to be the leading one, with the $q_3$ ordering also diverging but with smaller absolute magnitude. 
The increase of $\Lambda_c$ can be understood in terms of the Stoner criterion being fulfilled at a larger scale for larger $U$ or $J$.
At higher energy scales, the vHs are strongly smeared. This effect increases the importance of the $q_2$ ordering vector connecting two vHs points, leading to the emergence of the $q_2$-SDW phase.

When applying uniaxial strain, the parameter range where we find a SDW is increased.
This can be understood from the increase of the DOS at the Fermi level, which leads to a larger $\chi^{0}_{\text{PH}}$ and thereby a smaller interaction is required to fulfill the Stoner criterion.
Beyond this effect, straining does not affect the structures of the phases and the $q_2$-phase is still observable in the same parameter region, as the changes of the FS due to strain have counteracting effects: while in the $y$-direction the 
FS touches the vHs, it drifts further away from it in the $x$-direction.

Note that as we increase the strain beyond the Lifschitz transition, 
we do not find the SDW that is observed in experiments~\cite{grinenko_split_2021, li2022elastocaloric}.
The emergence of this phase has been understood as the removal of all curvature of the $\gamma$ sheet between the upper/lower vHs and the $X/X'$ points, which leads to strong nesting along this direction~\cite{PhysRevB.102.054506}.
We do not observe this phase at any investigated strain value. As was pointed out in Ref~\cite{PhysRevLett.130.026702}, strain suppresses quantum fluctuations preventing the ordered magnetic state, such that at high strain the phase transition emerges. Therefore, observing this transition in a diagrammatically unbiased approach is currently out of reach numerically, as it would require a full inclusion of the frequency dependence as well as multi-loop contributions~\cite{Hille_2020}. We hope to overcome this shortcoming in the future.

\begin{figure*}[!hbt]
    \centering
    \includegraphics[width = 0.78\linewidth]{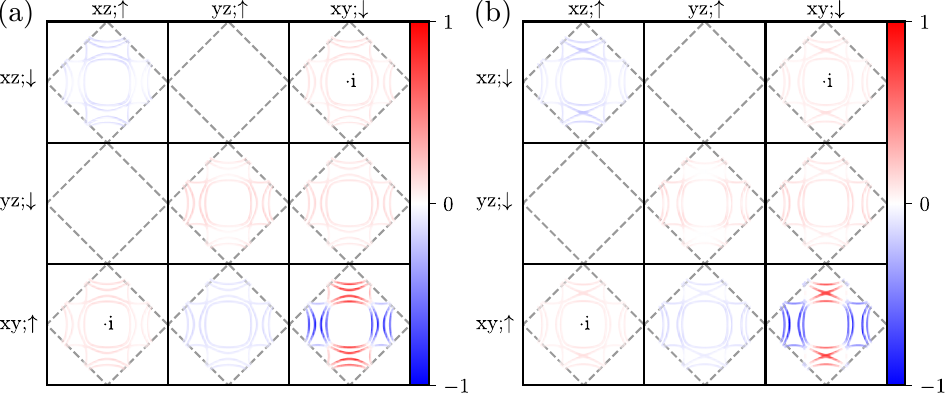}
    \caption{\textbf{Example of superconducting order parameters} found in Fig.~\ref{fig:phase_diag}, calculated from a linearized gap equation in orbital space.
    They correspond to $U = 1.1$~eV and $J = 0.22~U$. The $3\times3$ matrices represent the spin-orbital states of the paired electrons in the inter-pseudospin channel since the intra-pseudospin terms are vanishing.
    Each panel shows the momentum structure. We present in panel (a) and (b) the unstrained case with $\epsilon_{\rm{xx}} = 0$ and the optimally strained case with $\epsilon_{\rm{xx}}^{\text{vHs}}$ respectively.
    The SCOP transforms like the one-component B$_{\rm{1g}}$ (A$_{\rm{g}}$) irreducible representation of the D$_{\rm{4h}}$ (D$_{\rm{2h}}$) group which requires some components to be purely imaginary due to the transformation behavior of the spin-orbitals under rotational symmetries~\cite{gingras_superconductivity_2022}.
    Since the gap function is only known up to a prefactor, we rescale it from $-1$ to $1$, the sign and value is encoded in the colorbar. The first Brillouin-zone is marked by the grey square.
    }
    \label{fig:d-wave_state}
\end{figure*}

\paragraph*{Superconductivity. ---}
In the following, the superconducting phase is analyzed using a linearized gap equation on the FS.
As shown in Fig.~\ref{fig:d-wave_state}, we find a gap that transforms according to a B$_{\rm{1g}}$ for $\epsilon_{\rm{xx}} = 0$ (A$_{\rm{g}}$ 
for $\epsilon_{\rm{xx}}=\epsilon_{\rm{xx}}^{\text{vHs}}$) representation of the D$_{\rm{4h}}$ (D$_{\rm{2h}}$) point groups.
In the band basis, this state has a dominant overlap with the $d_{\rm{x}^2-\rm{y}^2}$  harmonic and its main weight stems from the $d_{\rm{xy}}$ orbital.
Such a type of superconductor has been observed in several other studies~\cite{PhysRevResearch.4.033011, romer_superconducting_2021, PhysRevB.106.134512, PhysRevB.106.054516, PhysRevB.106.L100501, romer_superconducting_2021, PhysRevB.102.054506, PhysRevLett.107.277003, gingras_superconducting_2019, gingras_superconductivity_2022}. 
Although spin-orbit mixing distributes pairing over the different FS sheets, 
the dominant $d_{\rm{xy}}$ orbital character that we find does lead to a larger pairing on 
the $\gamma$ sheet. It should be noted however that experiments indicate that the gap function is sizeable 
on all three Fermi surface sheets.This potential difficulty could be resolved by employing an analysis based on the combination of FRG and mean-field theory~\cite{PhysRevB.89.121116}.

The spectrum of the pair-pair susceptibility at $\Lambda_c$ contains the information of all possible subleading SCOPs.
By analyzing this spectral distribution, we find a clear separation of the eigenvalue of the $d_{\rm{x}^2-\rm{y}^2}$ superconducting state by at least one order of magnitude from all eigenvalues of other SCOPs, for all parameters investigated.
While this excludes any immediate degenerate state, no statement about the 
proximity of different symmetry states or individual critical temperatures can be drawn from FRG, 
because within this method the dominant instability is signalled by 
a divergent coupling and susceptibility ~\footnote{Again we stress that $\Lambda_c$ is not to be confused with the leading eigenvalue from an Eliashberg calculation $\lambda$. Most importantly, we have $\Lambda_c \propto T_{\rm{c}}$ instead of the exponential suppression from the gap equation.}.
However, from the hierarchy standpoint, we still can extract tendencies towards different orderings as discussed in App.~\ref{app:glue}.
This hierarchy reveals that the
$\rm{p}$-wave pairing state~\cite{PhysRevB.97.224522,PhysRevLett.122.027002,wang2013theory} 
is always clearly subleading by a large margin and both extended s-wave and $g$-wave are consistently closer to the $d$-wave.
We simulate the full 3D model at a selected point to check for consistency of our 2D simulations. The results of these 3D simulations are shown in App.~\ref{app:3D} and agree with the 2D results.

The superconducting phase is generated by a spin-fluctuation mechanism. 
The couplings $U$ and $J$ are crucial tuning knobs determining the onset of the phase and also control the transition to the neighbouring magnetic phase.
When increasing $U$, the transition to a SDW is understood from the underlying ladder-type diagrams diverging as soon as $U$ becomes larger than the critical value.
Below that critical $U$, the still strong spin-fluctuations can drive a superconducting instability.
However, increasing $J$ has a more complex effect since it affects two different physical processes, which we 
discuss in terms of two distinct couplings, 
$J_{\rm{ss}}$ and $J_{\rm{dd}}$ in Eq.~(\ref{eq:kanamori-ham}).
$J_{\rm{ss}}$ promotes spin-flip and pair-hopping processes, thus reducing the tendency to order magnetically while also increasing pair-correlations.
$J_{\rm{dd}}$ decreases inter-orbital density-density interactions, which reduces the inter-orbital repulsion between electrons on the same site.
We unravel which of the two effects is most relevant for a) superconductivity and b) the magnetic transition. This is achieved by varying the two quantities independently, first in a simple RPA calculation and then in a full FRG calculation.

For the simple RPA calculation, we calculate $\chi^0_{\text{PH}}$ at $\Lambda = 11.6$~meV.
We chose $U = 0.3\;$eV to circumvent the Stoner instability and vary $J_{\rm{ss}}$ and $J_{\rm{dd}}$ between $0.0U$ and $0.3U$ independently.
The dominant components of the bare susceptibility are presented in Fig.~\ref{fig:djj_vs_sjj} of App.~\ref{app:jddvjss}.
In general, we observe that varying $J_{\rm{ss}}$ has barely any effect on $\chi^{\text{RPA}}_{\text{PH}}$.
$J_{\rm{dd}}$, on the other hand, increases the inter-orbital components by a significant amount.
Therefore we expect $J_{\rm{ss}}$ to have a weaker impact on the superconducting transition. Physically this is expected since $J_{\rm{ss}}$ hampers the spin-fluctuations which are required to obtain an effective attraction required by the superconducting state.
To support this claim and understand better the underlying interference mechanism, we developed a simple 2-band toy model in App.~\ref{app:jddvjss}.

In the full FRG simulation, we verify these conclusions, i.e. increasing $J_{\rm{ss}}$ leads to a transition only at much larger values than the one for $J_{\rm{dd}}$. See Fig.~\ref{fig:FRGdjj_vs_sjj} of App.~\ref{app:jddvjss}.
Interestingly, $J_{\rm{ss}}$ will generate a stronger admixture of higher order angular momentum superconductivity hosted by the $d_{\rm{yz}}$ and $d_{\rm{xz}}$ orbitals. These are however still sub-leading to the $d_{\rm{x}^2-\rm{y}^2}$ superconducting state. 

\paragraph*{Influence of strain. ---}
\begin{figure}[!bht]
    \centering
    \includegraphics[width = \columnwidth]{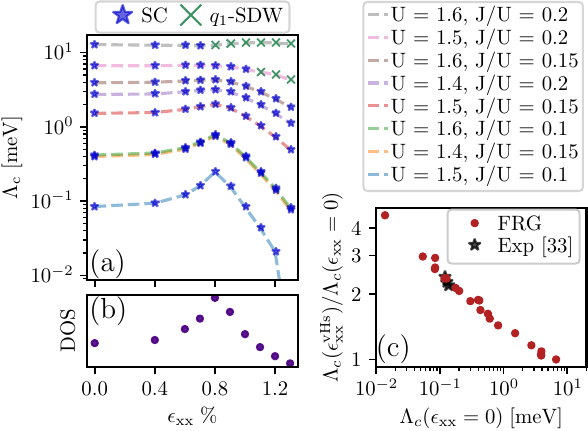}
    \caption{
    \textbf{Effect of strain $\epsilon_{\rm{xx}}$ on the critical scale $\Lambda_c$} for different values of $U$ and $J$ (upper left) and values of $U$ and $J$ for each line given in the upper right. DOS in the $d_{\rm{xy}}$ orbital depending on $\epsilon$ (lower left) and theoretically predicted enhancement of $\Lambda_c$ due to uniaxial strain as a function of its value for $\epsilon_{\rm{xx}} = 0$ (lower right).
    Each dotted line in the upper left plot corresponds to one $U-J$ combination given in the upper right.
    There is a clear correlation between $\Lambda_c(\epsilon_{\rm{xx}}=0)$ and the ratio of increase in $T_{\rm{c}}$ which can be seen in both the upper left and lower right panels. The experimental data points in the lower right plot are extracted from~\cite{steppke2017strong}.
    A proposed mechanism that explains this enhancement is detailed in the text.
    }
    \label{fig:crit_scale_vs_strain}
\end{figure}
Finally, we compare our results with experiments.
We do so by examining the effect of strain from $\epsilon_{\rm{xx}} = 0.0\%$ to $\epsilon_{\rm{xx}} = 1.3\%$ on the leading instability of different $(U,\ J)$ combinations. The general behavior of $T_{\rm{c}}$ is consistent with earlier studies~\cite{PhysRevB.102.054506, Liu_2017,PhysRevResearch.2.032055}, while the predicted phases partially differ.
The different critical scales $\Lambda_c$ can be interpreted as an estimate for $T_{\rm{c}}$ of the instability.
The results for all superconducting data points are summarized in Fig.~\ref{fig:crit_scale_vs_strain}.

For systems that start with a large initial critical scale at zero strain ($\Lambda_c(\epsilon_{\rm{xx}} = 0)$), no significant enhancement with respect to strain is found.
The enhancement of $T_{\rm{c}}$ is much larger when $\Lambda(\epsilon_{\rm{xx}} = 0)$ is smaller.
This effect can be understood by looking at the DOS: large energy scales, or large temperatures, correspond to smeared out features in the DOS.
Thus, the shift of the vHs due to strain is irrelevant since the vHs is not resolved, i.e.~the DOS at the Fermi level does not change under strain.
The lower $\Lambda_c$, the sharper the vHs will become. Therefore, its 
shift enhances the DOS at the FS more strongly which in turn leads to a larger increase of $T_{\rm{c}}$.
Thus, a lower $\Lambda_c(\epsilon_{\rm{xx}} = 0)$ yields an enhancement of $T_{\rm{c}}$ 
with $\epsilon_{\rm{xx}}$ 
which is both larger and taking place over a narrower range of strain.
Once the vHs has crossed the Fermi level, $T_{\rm{c}}$ is expected to go down again since we rapidly reduce the DOS at the Fermi level when straining further. Note that the nesting which generates the attractive interaction is not strongly asymmetric under strain, as can be seen in Fig.~\ref{fig:FS_and_DOS}. Therefore, it cannot explain the asymmetry of $\Lambda_C$ observed in Fig.~\ref{fig:crit_scale_vs_strain}. However, the density of states is asymmetric, see Fig.~\ref{fig:crit_scale_vs_strain}, giving rise to the observed asymmetry in $\Lambda_C$.

To compare our results to experiments, we evaluate $\Lambda_c(\epsilon_{\rm{xx}}^{\text{vHs}})/\Lambda_c(\epsilon_{\rm{xx}} = {0})$ and plot it versus $\Lambda_c(\epsilon_{\rm{xx}} = {0})$, hence measuring the increase of the critical scale depending on the initial one.
We extract the corresponding experimental values from Ref.~\citenum{steppke2017strong} by calculating the ratio of the maximal $T_{\rm{c}}$ and the $T_{\rm{c}}$ at $\epsilon_{\rm{xx}} = 0$.
These results are summarized in Fig.~\ref{fig:crit_scale_vs_strain}.
We observe that the experiments indeed fit to the data predicted by FRG and we can extract a line of $U$ and $J$ combinations along which the experiment is reproduced.
We find that the values on the line are around $U = 1.1, \ 1.4\;$eV and $J=0.143\,U, \ 0.1\,U$. 
Again, we emphasize that these should be considered as effective 
values valid within our FRG formalism.

\section*{\label{sec:conclusion}Discussion}
In summary, we studied SRO starting from 
a first-principles description of its electronic structure and using a diagrammatically unbiased FRG approach.
Using this framework, we investigated the influence of uniaxial strain as well 
as different contributions of Hund's coupling.
We identified that the inter-orbital interaction reduction due to the 
density-density term $J_{\rm{dd}}$ is the main driving force favoring superconducting order, which we found to be a pseudospin-singlet $d_{\rm{x}^2-\rm{y}^2}$.
Lastly, we showed that the experimental increase of $T_{\rm{c}}$ as a function of strain can be recovered on a quantitative level from FRG simulations and from a comparison to these experiments we extracted 
effective values of the interaction parameters.

Our results highlight the dominance
of a single $d_{\rm{x}^2-\rm{y}^2}$ SCOP that transforms like the B$_{\rm{1g}}$ representation (A$_{\rm{g}}$ under uniaxial strain). 
We note that, while this SCOP agrees with many experimental measurements, it cannot explain the evidence for two-components and time-reversal symmetry breaking.
From the experimentally observed behavior of the time-reversal symmetry condition, we can infer that a partner of our found SCOP is required to remain invariant under moving the vHs through the Fermi level.
This condition would be for example fulfilled by states with nodal lines along the $x$ direction~\cite{PhysRevB.104.054518, PhysRevB.108.014502} or odd-frequency superconductors~\cite{gingras_superconductivity_2022, gingras2023frequencydependent}.

An interesting direction for future studies would be to 
investigate the effect of interaction terms consistent with D$_{\rm{4h}}$ symmetry 
but breaking full cubic symmetry.
This could potentially influence the competition between different low energy 
orders~\cite{PhysRevLett.116.106402}.
There are also many potential routes towards a more accurate
investigation of the superconducting state.
First, even though SRO is nearly perfectly layered, including the third spatial dimension increases the number of allowed ordering types~\cite{Ramires_micros_perspective_2019}.
Secondly, taking into account the frequency-dependence of the 
gap function and self-energy would allow to gauge the relevance of the proposed odd-frequency state. 
Finally, an important advance would be to start the FRG flow from a 
correlated starting point \textit{i.e.}~from dressed quasi-particles and their effective interactions. 
This could for example be achieved by starting the FRG from a DMFT~\cite{PhysRevLett.112.196402} description of the normal state.


\section*{\label{sec:methods}Methods}
\paragraph*{Density functional Theory. ---}
We use density functional theory~\cite{kohn_self-consistent_1965, hohenberg_inhomogeneous_1964, kohn_density_1996} and the Quantum ESPRESSO DFT package~\cite{giannozzi2009quantum, zhang2017p} with the PBE exchange-correlation functional~\cite{Perdew:1996} to calculate the electronic structure.
Cell parameters and internal coordinates of the crystal structure in the $I4/mmm$ space group are relaxed in the conventional cell until all force components are smaller than 1~mRy/$a_0$ ($a_0$: Bohr radius) and all components of the stress tensor are smaller than 0.5~kbar, yielding a relaxed in-plane (out-of-plane) lattice constant of $a_0=3.878$~\AA ($c=12.900$~\AA).
To calculate the strained structures, we fix one in-plane lattice constant of the conventional cell to the strained value, $a_{\text{new}} = (1 - s) a_0$, and relax the two orthogonal cell parameters as well as the internal coordinates as described above.
After relaxation, we use the corresponding primitive unit cells containing one ruthenium atom each, i.e.~three $t_{2g}$ orbitals.
We use scalar-relativistic ultrasoft pseudopotentials from the GBRV library~\cite{Garrity:2014}, with the 4$s$ and 4$\rm{p}$ (2$s$) semicore states for both strontium and ruthenium (for oxygen) atoms included in the valence.
In the scalar-relativistic approximation, the spin-orbit coupling term is dropped~\cite{Takeda1978}.
The energy cutoffs for the wave functions and charge density are set to 60~Ry and 720~Ry, respectively.
We use a $12 \times 12 \times 12$ Monkhorst–Pack $k$-point grid to sample the Brillouin zone, and a smearing of 0.01~Ry utilizing the Methfessel–Paxton scheme.
To describe the low-energy physics we construct three ruthenium-centered $t_{2g}$-like maximally localized Wannier functions for each strained structure using Wannier90~\cite{mostofi2008wannier90,mostofi2014updated,pizzi2020wannier90}.
Spin-orbit coupling is included by first performing the DFT 
calculation without it 
and then adding a local SOC $\lambda_{\text{SOC}}=200$~meV to account for the correlation-induced enhancement over the DFT value.

\paragraph*{Functional renormalization group. ---}
In order to account for strong local electronic correlations in this multi-orbital system, we consider the Hubbard-Kanamori interaction Hamiltonian~\cite{kanamori1963electron}
\begin{align}
    \label{eq:kanamori-ham}
        \bm \hat{H}_{\text{int}} & = 
        \sum_{\rm{il}} U \hat{n}_{\rm{il}}^{\uparrow} \hat{n}_{\rm{il}}^{\downarrow} +
        \sum_{\rm{il_1}\neq \rm{l_2}} (U - 2J_{\rm{dd}}) \hat{n}_{\rm{il_1}}^{\uparrow} \hat{n}_{\rm{il_2}}^{\downarrow} \nonumber \\ &
        + \sum_{\rm{i}\sigma \rm{l_1} \neq \rm{l_2}} (U-3J_{\rm{dd}}) \hat{n}_{\rm{il_1}}^{\sigma} \hat{n}_{\rm{il_2}}^{\sigma}\\ &
        - \sum_{\rm{il_1} \neq \rm{l_2}} J_{\rm{ss}} \hat{c}^{\uparrow, \dag}_{\rm{il_1}} \hat{c}_{\rm{il_1} }^{\downarrow} \hat{c}^{\downarrow,\dag}_{\rm{il_2}}\hat{c}_{\rm{il_2}}^{\uparrow} + \sum_{\rm{il_1} \neq \rm{l_2}} J_{\rm{ss}} \hat{c}^{\uparrow,\dag}_{\rm{il_1}} \hat{c}^{\downarrow,\dag}_{\rm{il_1}} \hat{c}_{\rm{il_2}}^{\downarrow} \hat{c}_{\rm{il_2}}^{\uparrow} \nonumber
\end{align}
where $U$ is the intra-orbital on-site Coulomb repulsion, while $J_{\rm{dd}}$ ($J_{\rm{ss}}$) is the density-density (spin-flip and pair hoping) part of the Hund's coupling.
In the rotationally invariant formulation where $O(3)$ symmetry is satisfied, $J_{\rm{dd}} = J_{\rm{ss}} = J$. Although usually equal, the distinct effect of $J_{\rm{dd}}$ and $J_{\rm{ss}}$ when they are treated independently is discussed in Appendix~\ref{app:ph_susz}. Our calculations are performed in the regime of $U/W = 0.3-0.5$, with $W$ being the total bandwidth and $U$ being the Hubbard interaction.
We restrict ourselves to a situation with an on-site Hubbard-Kanamori interaction, as with this type of interaction, we reproduce the dominant spin-spin suszeptibilities peak position in comparision to the interacting spin-spin susceptibility measured in neutron scattering experiments~\cite{PhysRevLett.122.047004}, see App.~\ref{app:ph_susz}. Any interaction resulting in an analog peak structure should give the same physical results, as we argue in App.~\ref{app:glue}.

The strong electronic correlations emerging from the Hubbard-Kanamori interactions are incorporated to the non-interacting downfolded systems using the functional renormalization group (FRG)~\cite{metzner_functional_2012,Dupuis_FRG_review}.
FRG is technically an exact method to calculate the effective action functional of a given quantum action.
It does so by introducing a scale-dependent cutoff (here we use a sharp energy cutoff) in the non-interacting propagator of the system.
By taking derivatives with respect to this cutoff, one generates an infinite hierarchy of flow equations.
In practice, this hierarchy must be truncated to become numerically tractable, making the method pertubatively motivated.

In this work, we employ the standard level-2 truncation, neglecting all three and more particle vertices.
The validity of this approximation in the weak-to-intermediate coupling regime can be motivated by a power-counting argument to prove the RG-irrelevance of higher order terms~\cite{metzner_functional_2012}.
Furthermore, we neglect the frequency dependence of the interaction, again motivated by the power counting argument, and the self-energy.
This approach was applied to various systems including SRO~\cite{PhysRevLett.122.027002, Liu_2017, PhysRevB.101.064507, PhysRevB.97.224522, scherer2022chiral,Klebl_2023,PhysRevB.106.125141, wang2013theory,PhysRevLett.111.097001, PhysRevLett.110.126405, PhysRevB.107.125115, PhysRevB.86.020507, Hauck_2021, PhysRevB.102.195108, PhysRevB.102.085109, PhysRevB.96.205155, PhysRevB.99.245140} and can be viewed as an diagrammatically unbiased extension of the random phase approximation.

In practice, we solve the flow equations from an energy scale much larger than the bandwidth and then integrate towards lower energies until we hit a divergence in one of the three diagrammatic channels labelled the particle-particle (PP), particle-hole (PH) and crossed particle-hole ($\overline{\text{PH}}$) channels.
A divergence is associated to a phase transition as the corresponding susceptibility also diverges.
Information of the ordering type can be extracted from the susceptibilities as well as linearized gap equations~\cite{Beyer_2022}.

We employ the truncated unity approximation which allows us to reduce the memory required computationally~\cite{SMFRG-hussalm, lichtenstein_high-performance_2017, Hauck_2022}. 

For the FRG simulations, we use the TU$^2$FRG code~\cite{Hauck_2022}. For convergence, we include all form-factors up to a distance of $8.2$~\AA, which amounts to a total number of 75 basis functions per orbital in the unit cell, i.e.~all lattice harmonics up to the fifth (sixth) are included (depending on whether the constant is counted as harmonic or not).
We checked for convergence by increasing the number of form-factors included near the phase transition between the magnetic and the superconducting phase for a few data points in the phase diagram.
The simulations are performed on a $36\times 36$ momentum-mesh in the $x-y$ plane for the vertex function.
The loop integration is performed using a FFT approach and an additional refinement of $45\times 45$ is employed to achieve higher energy resolution.
The results of the integration do not differ upon changing the resolution of the loop integration.

By using an enhanced value of the SOC, 
the effects of local interactions on SOC are already included on the single-particle level of the calculations.
We do not suffer from double counting at that level since we neglect the flow of the self-energy.
As a consequence however, our calculation does not take into account the renormalized effective mass of quasi-particles.

\paragraph*{Acknowledgments. ---}
JBP thanks Prof.~E.~Pavarini, Friedrich Krien, Lennart Klebl and Jacob Beyer for helpful discussion. 
The authors gratefully acknowledge the computing time granted through JARA on the supercomputer JURECA~\cite{JUWELS} at Forschungszentrum Jülich.
JBP and DMK are supported by the Deutsche Forschungsgemeinschaft (DFG,
German Research Foundation) under RTG 1995, within the Priority Program
SPP 2244 ``2DMP'' --- 443273985 and under Germany's Excellence Strategy - Cluster of
Excellence Matter and Light for Quantum Computing (ML4Q) EXC 2004/1 -
390534769. The Flatiron Institute is a division of the Simons Foundation.  

\paragraph*{Competing Interest. ---}
The Authors declare no Competing Financial or Non-Financial Interests.

\paragraph*{Data availability. ---}
All simulation data is available upon reasonable request. 

\paragraph*{Code availability. ---}
The codes used for simulation are open source packages and as such freely available.

\paragraph*{Author Contributions. ---}
SB performed the DFT simulations. JBP performed the FRG simulations. JBP, AG and OG analyzed the results. All authors contributed in writing the Manuscript.

\bibliography{Sr2RuO4}

\appendix
\section{Low-energy tight-binding model}
\label{app:dft}
\begin{figure}[!hbt]
    \centering
    \includegraphics[width = 0.38\columnwidth]{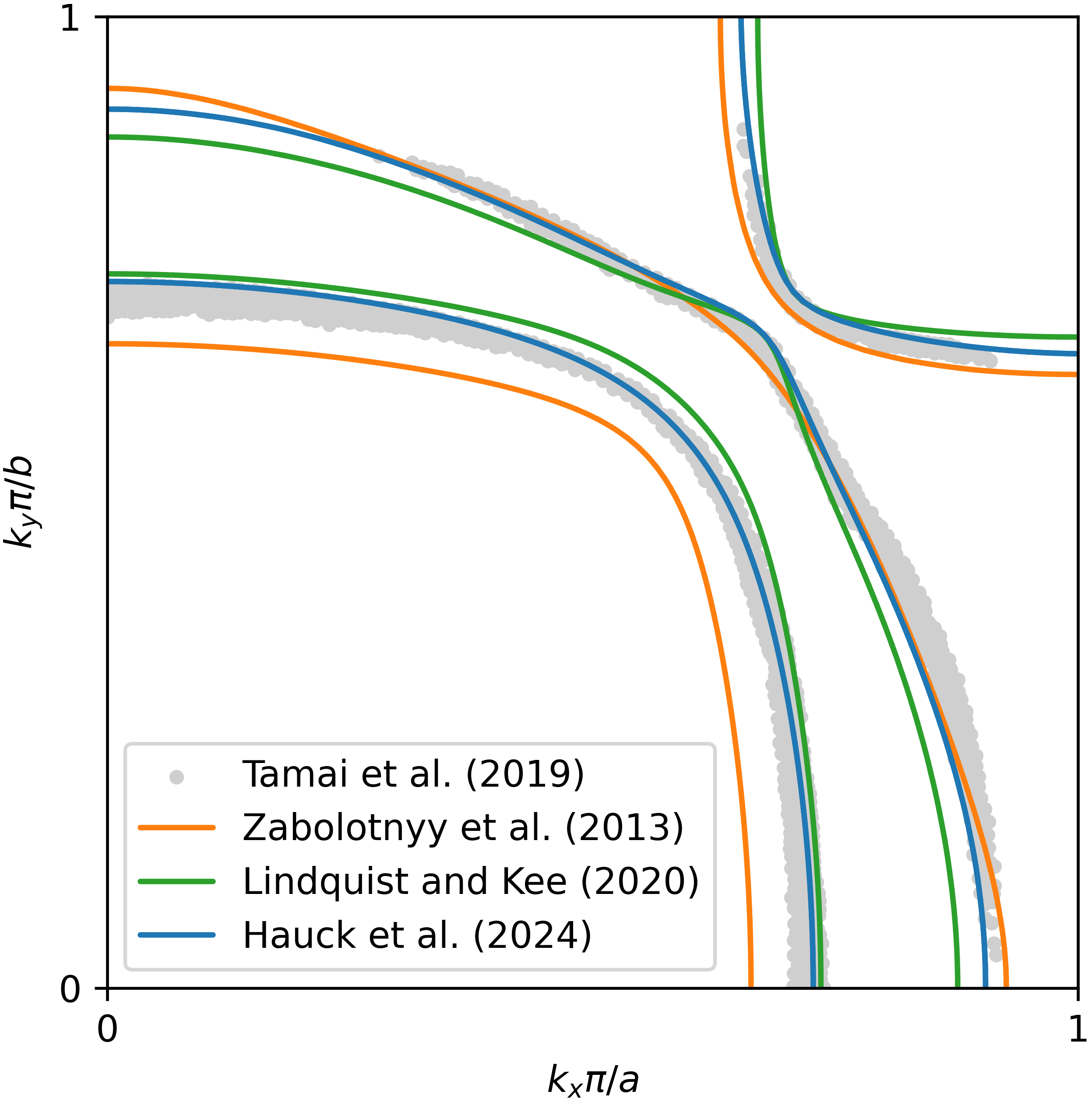}\hfill
    \includegraphics[width = 0.59\columnwidth]{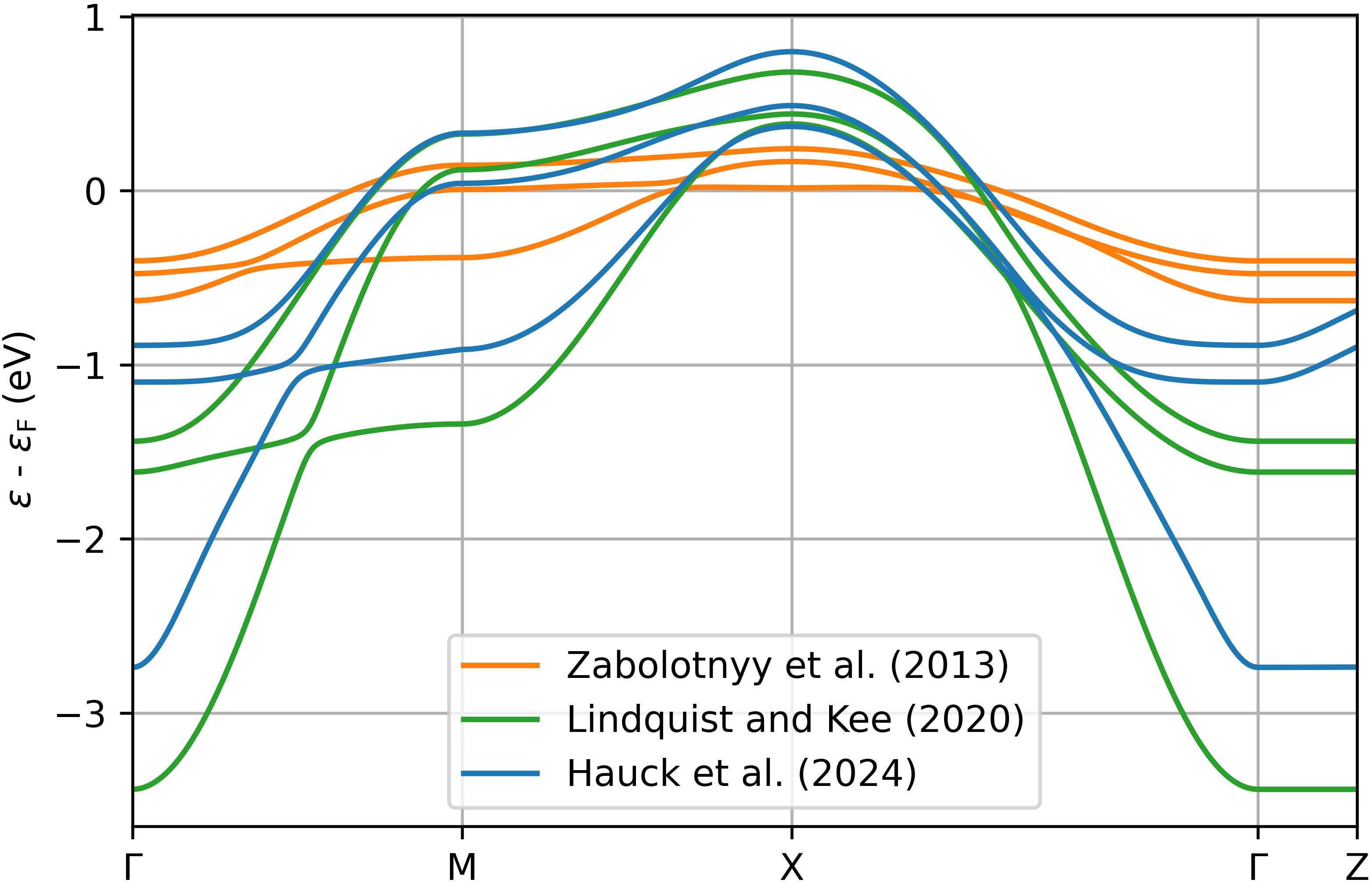}
    \caption{\textbf{Comparison of non-interacting models.} Comparison of the Fermi-surface (left) and the bandstructure along the high-symmetry path (right) between this work, Zabolotnyy et. al~\cite{Zabolotnyy_2013}, Lindquist and Kee~\cite{PhysRevResearch.2.032055} and the ARPES data from Ref.~\cite{tamai_high-resolution_2019}. Note that we neglect the $t_5$ hopping in Ref.~\cite{PhysRevResearch.2.032055} as it was unclear to what term it corresponds.}
    \label{fig:FS_compare}
\end{figure}

The tight-binding Hamiltonian used in this work, based on DFT and a correlation-enhanced SOC, reproduces the experimentally observed Fermi-surface of Ref.~\cite{tamai_high-resolution_2019} as shown in Fig.~\ref{fig:FS_compare}.
Furthermore, our DFT results are consistent with prior DFT results~\cite{gingras_superconducting_2019, moon2023effects} within the different spin orbit couplings considered.
Larger differences arise in comparison with the Fermi surface in Refs.~\cite{Zabolotnyy_2013,PhysRevResearch.2.032055}, see Fig.~\ref{fig:FS_compare}. These works extracted a tight-binding model from fits to the Fermi surface measured in ARPES. However, recent advances in ARPES techniques allow a more precise measurement of the Fermi surface, explaining the difference to Ref.~\cite{Zabolotnyy_2013}. On the other hand Ref~\cite{PhysRevResearch.2.032055} fits to Tamai et. al~\cite{tamai_high-resolution_2019}, thereby obtaining a similar Fermi surface to the one presented here, however, with a much larger bandwidth. Since the fit is optimized for the Fermi surface only, the bandwidths of these models need not necessarily agree and are shown here for illustrative purposes only, in particular since some of the fits represent quasi-particle bands.
A comparison of the Fermi surface and bandstructure of this publication and various ARPES fits is shown in Fig.~\ref{fig:FS_compare}.

\section{Particle-hole susceptibility}
\label{app:ph_susz}

\begin{figure}[!hbt]
    \centering
    \includegraphics[width = \columnwidth]{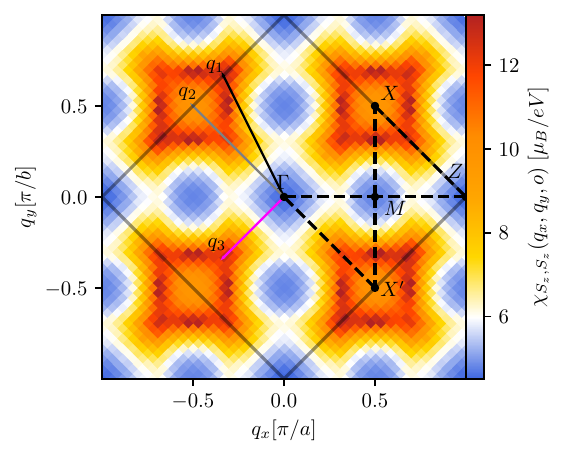}
    \caption{\textbf{Interacting particle-hole susceptibility at $\Lambda = 464$~K obtained with FRG} at the extracted experimental interaction parameters $U = 1.1$~eV and $J=0.143~U$.
    The color scale is adapted to the one of Ref.~\cite{Strand_2019}.
    We again mark the first BZ as a black square and draw the $k_z = 0$ irreducible path as dotted lines.}
    \label{fig:intph_susz}
\end{figure}

In the main text, we explained how we can extract the value of effective parameters of the interaction suitable in FRG, here we found $U = 1.1$~eV and $J = 0.143~U$.
With these extracted values, we compute the interacting particle-hole susceptibility $\chi_{\text{PH}}$ at $T = 464$~K, shown on a two-dimensional momentum-grid in Fig.~\ref{fig:intph_susz}.
We compare with $\chi_{\text{PH}}$ obtained by solving the Bethe-Salpeter equation with a vertex extracted using DMFT~\cite{Strand_2019}.
FRG clearly overestimates the correlations at the $X$-point and on the connection line between $X$ and $M$, similarly to what is obtained using the random phase approximation~\cite{gingras_superconducting_2019, gingras_superconductivity_2022}.
Beyond that, FRG reproduces roughly the same shape and structure of the susceptibility as DMFT, but cannot reproduce its shifting of the peaks~\cite{Strand_2019, moon2023effects}.
As discussed in App.~\ref{app:glue}, correcting for this overestimation in the effective interaction might influence the critical energy scales, but is not expected to drastically alter the hierarchy of the different order parameters.
Most critically, we show analytically that the leading SDW peak observed in experiments~\cite{PhysRevLett.122.047004} will also lead to an attractive interaction in the singlet channel, see Appendix~\ref{app:glue}.

\section{Subleading instablilities}
\label{app:svt}
\begin{figure}[!hbt]
    \centering
    \includegraphics[width = 0.48\columnwidth]{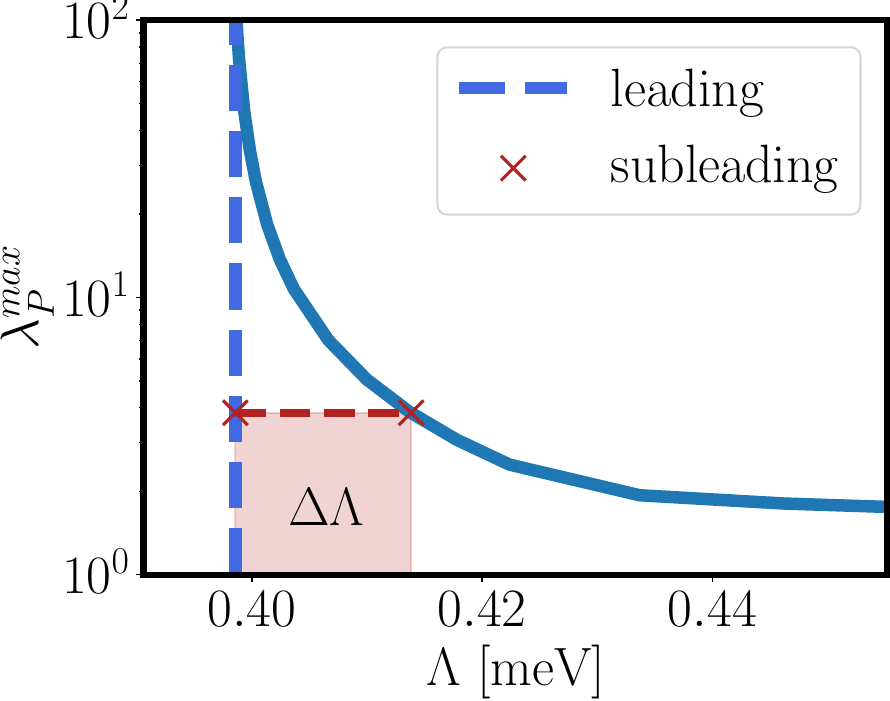}
    \includegraphics[width = 0.48\columnwidth]{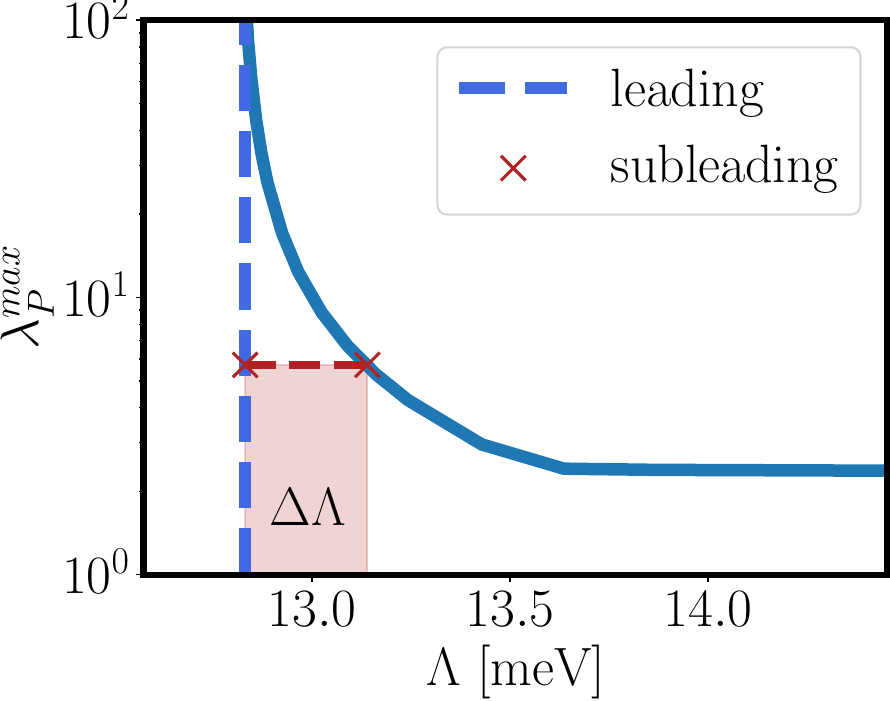}
    \caption{\textbf{Estimating the separation between the critical scales for the leading and subleading instability} at $(U,J/U) = (1.3\; \text{eV},0.1625)$ (left) and $(U,J/U)  = (1.5\;\text{eV},0.225)$ (right). The dashed vertical line marks the critical scale at the divergence. The value of the first subleading coupling is then extracted and tracked back to where the leading coupling had this value, this is shown in red. This gives us a lower bound on the separation of the divergent couplings. The separation is highlighted as the red area under the blue curve. This analysis gives an lower bound for the separation of the order of $2$ to $3\%$ under the assumption that the subleading coupling diverges at the same rate and with the same shape as the leading one.}
    \label{fig:track}
\end{figure}

To estimate the minimal separation between the divergence of the leading and the subleading eigenvalue we track at which scale the subleading instability's magnitude is the same as the leading instabilities one, see Fig.~\ref{fig:track}. This estimate assumes that the subleading and leading divergences are behaving identically. As such it is an approximate lower bound for the separation. We find it to be consistently to be in the order of a few percent.

Prior FRG studies~\cite{PhysRevB.97.224522,PhysRevLett.122.027002,wang2013theory} predicted a leading $\rm{p}$-wave divergence, which in our calculation is far subleading with at least 20 other even-parity eigenvectors between the leading and first odd-parity solution.
Thus we can safely conclude that odd-parity is far from dominant. The difference in results to prior FRG studies can be traced back to a more refined non-interacting model. I.e.~capturing the position of the largest peak observed in neutron diffraction is central to capture the leading contributions to the pairing in a spin fluctuation picture.

\section{Effects of \texorpdfstring{$J_{\rm{dd}}$ and $J_{\rm{ss}}$}{Jdd and Jss}}
\label{app:jddvjss}

\begin{figure}[!hbt]
    \centering
    \includegraphics[width = \columnwidth]{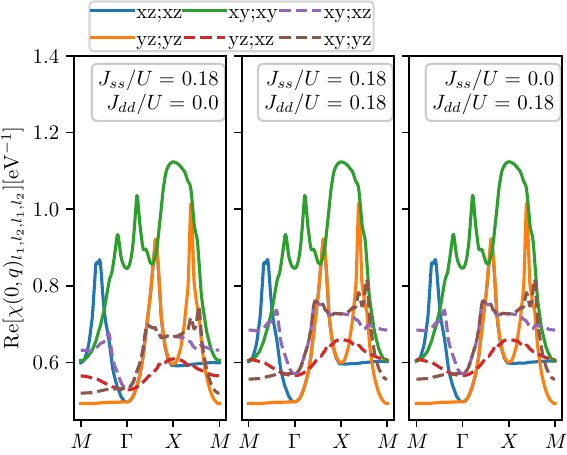}
    \caption{\textbf{Effects of $J_{\rm{dd}}$ and $J_{\rm{ss}}$ on the interacting susceptibility} obtained from RPA at $U=0.3$~eV and $\Lambda=10$~meV.
    }
    \label{fig:djj_vs_sjj}
\end{figure}
\begin{figure}[!hbt]
    \centering
    \includegraphics[width = \columnwidth]{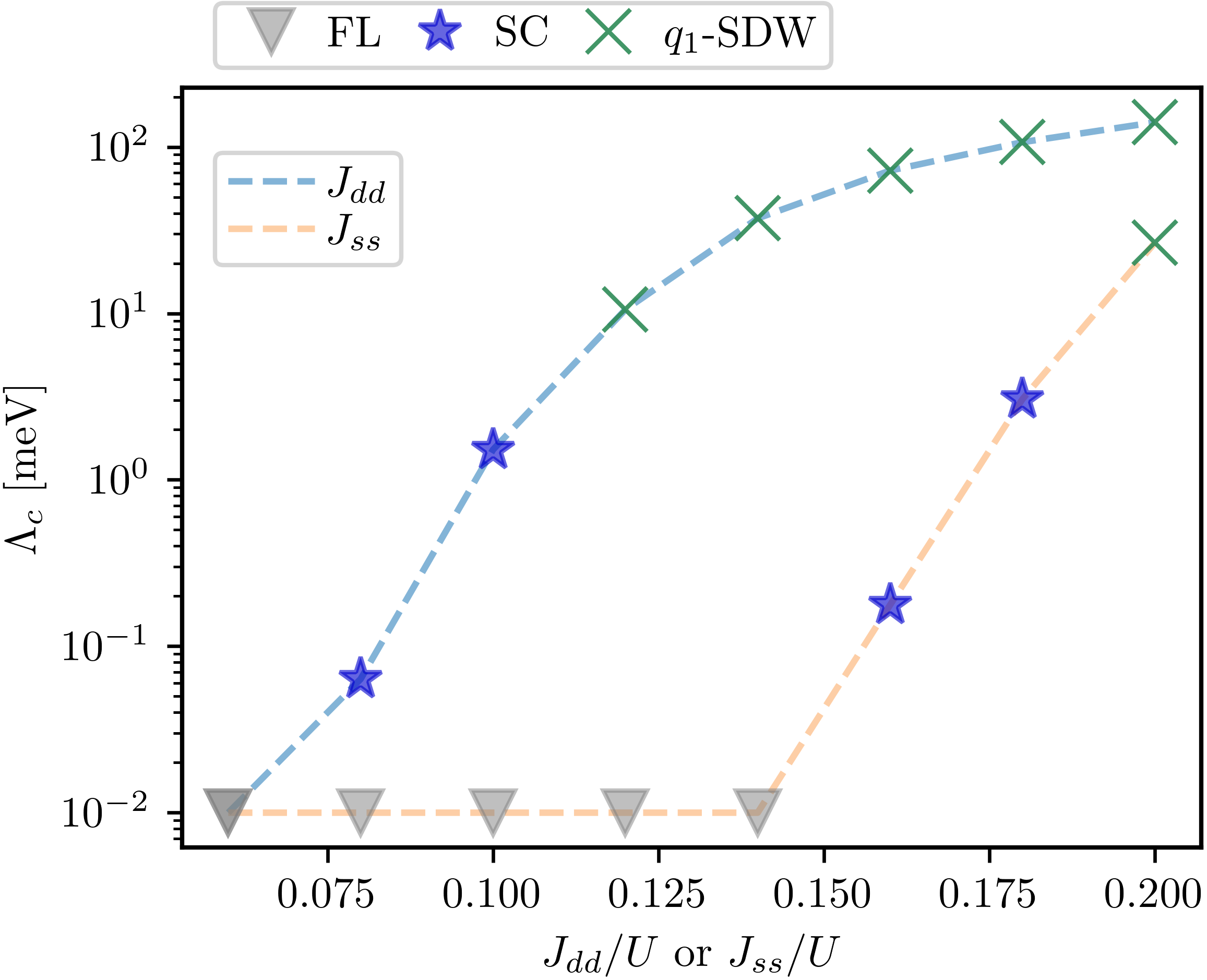}
    \caption{\textbf{Effects of $J_{\rm{dd}}$ and $J_{\rm{ss}}$ on the FRG flow results.} We observe that $J_{\rm{dd}}$ drives the transition to an ordered state quicker than $J_{\rm{ss}}$ indicating that for the underlying interaction mechanism it is beneficial to have smaller inter-orbital coupling.
    }
    \label{fig:FRGdjj_vs_sjj}
\end{figure}

In the following, we analyze analytically the effects of $J_{\rm{ss}}$ and $J_{\rm{dd}}$ in a simplified $SU(2)$ symmetric two-orbital model. This model is oversimplified so the results are not directly applicable to SRO, however it does help to understand the way in which the different contributions can influence the low energy physics. The non-interacting model is assumed to be given by
\begin{equation}
\hat{H}_0 = \sum_l \sum_{ij} t_{ij}(c^{\dagger}_{li}c_{lj}+c^{\dagger}_{lj}c_{li})
\end{equation}
where $l$ is an orbital and $i, j$ are sites. 
We first examine the magnetic channel susceptibility which is given in the two-particle basis $|l_1\rangle \otimes |l_2\rangle$.
The components of the bare particle-hole susceptibility are given by 
\begin{equation}
    [\chi^{0}_{\text{PH}}(Q)]^{l_1l_2l_3l_4}_{KK'} \propto G^{l_1l_3}_{K+Q} G^{l_4l_2}_K \delta_{KK'}
\end{equation}
where $G$ is the one-particle Green's function and $K$, $K'$ and $Q$ are four-momenta~\cite{bickers_self-consistent_2004, gingras_superconductivity_2022}.
From this expression, we find that the bare particle-hole susceptibility is diagonal in orbital space.
Presuming the two orbitals to be degenerate, we write
\begin{equation}
    \chi^0_{\text{PH}} = \left(
    \begin{array}{cccc}
        \chi_{1} & 0 & 0 & 0\\
        0 & \chi_{2} & 0 & 0\\
        0 & 0 & \chi_{2} & 0\\
        0 & 0 & 0 & \chi_{1}
    \end{array}
    \right).
\end{equation}

In RPA, the irreducible vertex in the particle-hole channel is the anti-symmetric static and local Coulomb tensor $\Gamma_{\text{PH}} = \Lambda_{\text{PH}}$.
We spin-diagonalize it and take the magnetic channel such that $\Lambda_{\text{PH}}^m \equiv \Lambda_{\text{PH}}^{\uparrow\uparrow\uparrow\uparrow} - \Lambda_{\text{PH}}^{\uparrow\uparrow\downarrow\downarrow}$. It describes the irreducible set of particle-hole interactions that generates spin-fluctuations.
As shown in the Appendix~C of Ref.~\citenum{gingras_superconductivity_2022}, it can be found, up to a minus sign due to conventions, as
\begin{equation}
    \Lambda_{\text{PH}}^m = \left(
    \begin{array}{cccc}
        U & 0 & 0 & J_{ph} \\
        0 & U-2J_{\rm{dd}} & J_{sf} & 0 \\
        0 & J_{sf} & U-2J_{\rm{dd}} & 0 \\
        J_{ph} & 0 & 0 & U
    \end{array} \right).
\end{equation}
The parameters are the same as those characterizing the Hamiltonian of Eq.~(\ref{eq:kanamori-ham}), in particular with the pair-hopping and spin-flip terms characterized by $J_{ph}$ and $J_{sf}$ respectively.
Consequently, the eigenvalues of $\Lambda_{\text{PH}}^m \chi^0_{\text{PH}}$ which quantify the amount of spin-fluctuations are
\begin{equation}
   \lambda =  \chi_1(U \pm J_{ph}), \ \chi_2(U-2J_{\rm{dd}} \pm J_{sf})
\end{equation}
with eigenvectors
\begin{equation}
    \left( \begin{array}{c} 1 \\ 0 \\ 0 \\ \pm 1 \end{array} \right)
    \quad \text{and} \quad
    \left( \begin{array}{c} 0 \\ 1 \\ \pm 1 \\ 0 \end{array} \right).
\end{equation}

In other words, a pair on $l_1$ can constructively or destructively interfere with a pair on $l_2$ via $J_{ph}$, which enhances the intra-orbital components of the susceptibility.
On the other hand, $J_{sf}$ enhances the inter-orbital components.
On the other hand, $J_{\rm{dd}}$ acts in the same way as spin flips, affecting only the inter-orbital components of the interacting susceptibility. However it has twice the magnitude.
Which terms play what role also crucially depends on the signs of $\chi_1$ and $\chi_2$; 

The corresponding numerical experiments are visualized in Fig.~\ref{fig:djj_vs_sjj}.
They quantify the effects of the two different couplings and help to gauge their importance for the phase diagram.

\section{Driving fluctuations analysis}
\label{app:glue}
Here, we want to construct a more in depth understanding of which spin-fluctuations are responsible for which pairing instabilities.
To this end, we focus on the $d_{\rm{xy}}$ orbital again assuming $SU(2)$ symmetry. Therefore we restrict the analysis to the spin-singlet sector. The spin-triplet can be obtained by employing the crossing relations. Furthermore, we assume an attractive interaction obtained from RPA like spin-fluctuation $C(q)$, i.e.~without fermionic momentum dependencies.
The starting point for this discussion is the linearized gap equation, which can be rewritten as
\begin{align}
    \lambda \Delta_{o_1,o_2}( k) = & \;\Gamma^P_{o_1,o_2;o_3',o_4'}( q =  0,  k, k') \nonumber \\
    & \times \chi_{o_3',o_4';o_3,o_4}( k') \Delta_{o_3,o_4}( k')
\end{align}
where summation over repeated indices is implicit.
We arrive to a simpler picture by transforming this equation to real-space, introducing the lattice vectors $ b$. We write
\begin{equation}
    \Delta_{o_1,o_2}( b) = e^{-i k \cdot  b} \Delta_{o_1,o_2}( k),
\end{equation}
which leads us to
\begin{align}
    \lambda \Delta_{o_1,o_2}( b) = & \;\Gamma^P_{o_1,o_2;o_3',o_4'}( b, b') \nonumber \\
    & \times \chi_{o_3',o_4';o_3,o_4}( b', b'') \Delta_{o_3,o_4}( b'').
\end{align}

Now, we are mainly interested in nearest neighbor superconductivity, thus restricting the allowed $ b$ to nearest neighbor form-factors.
The two central quantities in the linearized gap equation are the particle-particle loop $\chi_{o_3',o_4';o_3,o_4}( b', b'')$ and the P-channel vertex $\Gamma^P_{o_1,o_2;o_3',o_4'}( b, b')$, which in the following we construct for a specific choice of a symmetrized basis for the $d_{\rm{x}^2-\rm{y}^2}$ and the extended $s$-wave SCOPs for a purely spin-fluctuation interaction $C( q_C =  k+ k'- q_P)$. For this we introduce the symmetrized form-factors denoted by $f_{ b}( k)$:
\begin{widetext}
\begin{align}
    \Gamma^P( q_P =  0,  b, b) 
                    &= \frac{1}{4} \int d k \, d k'\, d r \; (\cos(k_x)\pm \cos(k_y))(\cos(k'_x)\pm \cos(k'_y))e^{i r( k +  k')}C( r) \nonumber \\
                    &= \frac{1}{4}\left[C(r_x = 1, r_y = 0) + C(r_x = -1, r_y = 0) + C(r_x = 0, r_y = 1) + C(r_x = 0, r_y = -1) \right].
\end{align}
\end{widetext}
From this we observe immediately two things: First, a pure spin-spin interaction does not differentiate between extended $s$-wave or $d_{\rm{x}^2-\rm{y}^2}$ (or $p_x$ and $p_y$).
It merely gives a numerical prefactor to be put into the gap equation.
Secondly, that prefactor is just dependent on the value of $C( r)$ for $ r$ on the nearest neighbors; i.e.~as long as the Fourier transform of C is attractive on the nearest neighbor, we generate superconductivity whose symmetry is determined by the particle-particle loops fermionic argument only. 
\begin{equation}
    C(r) = \int d q\; e^{i q r} C( q)
\end{equation}
For now, lets assume that $C(q)$ (again we focus on the $d_{\rm{xy}}$ orbital) is consisting of $\delta$-peaks at $q_2$ or $q_1/q_3$ with unit weight, for simplicity we furthermore go back to a square lattice unit cell.
This allows us to calculate the Fourier transformation $C(r)$ analytically.
We focus on $C(r_x = 1, r_y = 0)$, since all other terms can be understood via symmetries.
We find $C(r_x = 1, r_y = 0;q_2) = -1$  and $C(r_x = 1, r_y = 0;q_1) = e^{i2\pi/3}$, meaning that $\Gamma^P(q_2) = -1$ and $\Gamma^P(q_1) = -0.5$ on the nearest neighbor shell.
Thus the overestimation of correlations at $X$ amplifies tendency towards superconductivity, but not towards one specific type.
In more general terms, any transfer momentum of the form $q_i = (q,q)$ will generate attraction in the even channel as long as $rq_i > \pi/2$.
Below this threshold, the Fourier transformation of $C( q)$ will become positive and thus generate attractive interactions in the odd-channel. Thus also the predicted leading momentum transfer found in DMFT should generate attraction for a singlet state.

\begin{figure}[!hbt]
    \centering
    \includegraphics[width = \columnwidth]{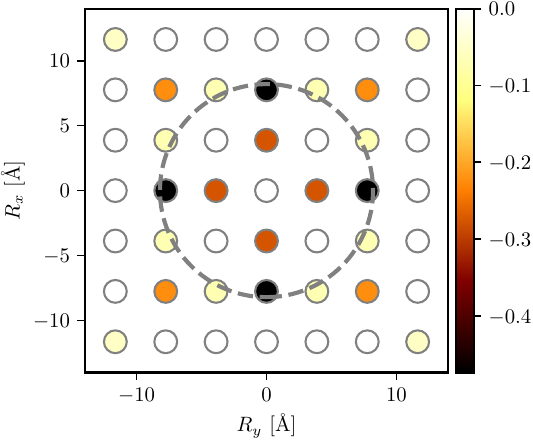}
    \caption{\textbf{Most attractive values of the projection of a spin-spin
    vertex to the pairing channel in the $x$-$y$ plane.} For this, we calculate an RPA-like flow for the C-channel up to a scale of $\Lambda = 1e^{-3}$. We plot here the real part of $\Gamma^P(b, b)$ calculated from the dominant eigenvalue of the $C$ channel at each $k$-point. To incorporate the two leading contributions, we require a from-factor cutoff of $7.8$\AA~in the full FRG calculation, visually indicated by a circle with radius $8.2$\AA~to emphasize that the largest attraction value is included.}
    \label{fig:where_att}
\end{figure}

To make again a connection to the model at hand, we analyse the attraction induced by an spin-fluctuation vertex for the full 3D model from RPA-flow. We present the attraction values generated by the leading eigenvalues of the interaction at each $q$ in the $xy$ plane, see Fig.~\ref{fig:where_att}.

\begin{figure*}[!hbt]
    \centering
    \includegraphics[width = 0.39\linewidth]{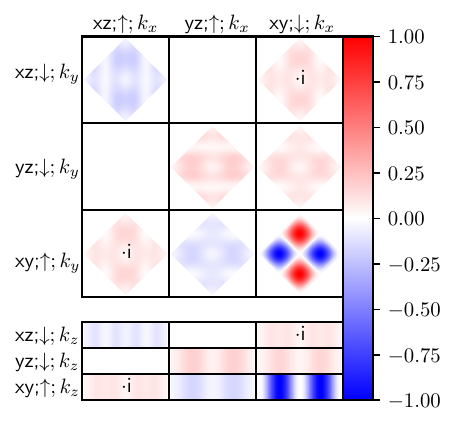}
    \includegraphics[width = 0.39\linewidth]{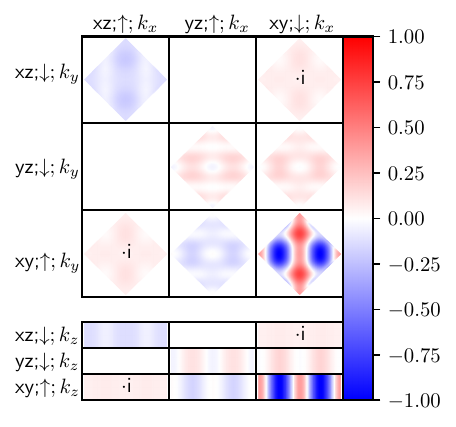}
    \caption{\textbf{2D cuts through the leading eigenvector of the pairing vertex from TUFRG} in $k_xk_y$ (upper panels) and $k_xk_z$ (lower panels) in spin-orbital-momentum representation for the unstrained (left) and $0.8\%$ strained (right) case. We find identical results to our 2D calculation.}
    \label{fig:3Dgap}
\end{figure*}

\begin{figure*}[!hbt]
    \centering
    \includegraphics[width = 0.39\linewidth]{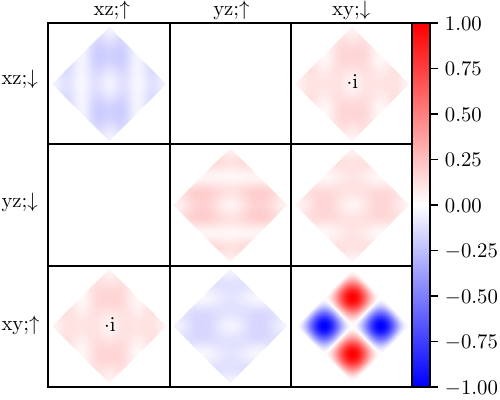}
    \includegraphics[width = 0.39\linewidth]{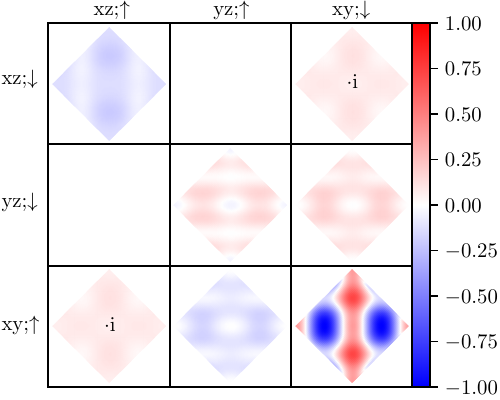}
    \caption{\textbf{Gap functions presented in Fig.~\ref{fig:d-wave_state} calculated on the full BZ}, for better comparability with Fig.~\ref{fig:3Dgap}}
    \label{fig:2DGap}
\end{figure*}

\section{Comparison to 3D results}
\label{app:3D}
To ensure that the results from our quasi 2D simulation are 
applicable to the realistic material, we check for consistency of our results at a few selected points between 2D and 3D calculations. In full 3D calculations more order 
parameters are allowed to be non zero~\cite{suh_stabilizing_2020,PhysRevResearch.4.023060,Ramires_micros_perspective_2019, Kaba_2019}, therefore besides the convergence check of our calculations this is also a consistency check of that the quasi-2D model indeed contains the leading instabilities of the 3D model. 
For this we simulate a $20^3$ grid for 
the vertex function with an additional refinement of $21^3$ for the
momentum integration in the loop. First, we estimate the most relevant
form-factor contributions. For this, we calculate $C_{o_1,o_3}({q})$ and 
apply what has been discussed in section~\ref{app:glue}, revealing that 
the most attractive contributions are the $x$ and $y$ direction nearest 
bond neighbor bonds as in our 2D simulations and their higher harmonics, see Fig.~\ref{fig:where_att}.

Thus, we include the first and second harmonic of this contribution in our calculation - i.e.~all neighbors within a distance of $7.8$\AA. The $q = 0$ SCOP results for $U=1.5$eV, $J=0.15U$ and $\epsilon_{\rm{xx}} = 0.0\%,\,0.8\%$ are summarized in Fig.~\ref{fig:3Dgap}.

Our results presented, indicate that the order parameter is indeed purely 2D, thus our 2D simulations capture the main aspects of the real material. Further the critical scale changes slightly from $3.15(3.69)$ meV in 3D to $2.48(2.80)$ meV in 2D with $0.0(0.8)\%$ uniaxial strain applied.
Two remarks have to be made: The nearly layered structure gives rise to sub-leading pair-density wave contributions with the same fermionic momentum dependence as presented in Fig.~\ref{fig:3Dgap} and finite transfer momentum of the form $q = (0,0,q_z)$. These are however subleading to the $q_z = 0$ SCOP. The existence of these competing pair-density waves can be understood from a simple 2D toy model - assume a two band system which is non-dispersive in one direction:
\begin{equation}
    H_{o_1,o_2}(k_x,k_y) = H_{o_1,o_2}(k_x)\;,
\end{equation}
this immediately implies that both the particle-hole and particle-particle loop have to be independent of $k_y$. Thus the spin-spin vertex is of the form
\begin{equation}
    V_{o_1,o_2}(k_x,k_y) = V_{o_1,o_2}(k_x)
\end{equation}
and the linearized gap equation also being independent of $k_y$. This implies that in the simple picture, all gaps of the form $\Delta_{o_1,o_2}(k_x,q_y)$ have equal eigenvalue $\lambda$ independent of $q_y$ explaining why we observe these pair-density waves in the nearly layered system.

\end{document}